\documentclass[pre,twocolumn,preprintnumbers]{revtex4} 
\usepackage{graphicx} 
\usepackage{dcolumn} 
\usepackage{bm} 
\usepackage{amsmath}

\usepackage[dvips]{color}

\usepackage{epsfig}

\begin{document}
\title{Time-reversal in dynamically-tuned zero-gap periodic systems} 
\author{Yonatan Sivan and John B. Pendry} \affiliation{The Blackett Laboratory,
Department of Physics, Imperial College London, London SW72AZ}
\email{ysivan@imperial.ac.uk}

\begin{abstract}
We show that short pulses propagating in zero-gap periodic systems can be reversed with $100\%$ efficiency by using weak non-adiabatic tuning of the wave velocity at time-scales that can be much slower than the period. Unlike previous schemes, we demonstrate reversal of {\em broadband} (few cycle) pulses with simple structures. Our scheme may thus open the way to time-reversal in a variety of systems for which it was not accessible before.
\end{abstract}


\maketitle

Wave propagation in periodic systems has been the focus of countless studies in several branches of physics. The most prominent of these systems are crystalline solids and their artificial analogues, known as photonic crystals (PhCs)~\cite{Joannopoulos-book}, phononic crystals~\cite{phononic_crystals} etc.. The study of such systems is almost exclusively associated with the forbidden gaps, which are energy/frequency regimes where waves cannot propagate. These gaps are crucial to understanding the electronic and optical properties of solid-state systems; their artificial analogues have a variety of possible applications, ranging from low-threshold lasers, light/sound guiding, filtering and switching~\cite{Joannopoulos-book} to medical ultrasound and nondestructive testing~\cite{phononic_crystals}.

Although the occurrence of bandgaps is generic to periodic systems, in some special cases, the gaps can have a {\em zero} width. Similarly, some materials and structures support two bands that cross symmetrically, {\em without} forming a gap. One of the most notable of those is graphene~\cite{graphene}, or its photonic crystal analogue~\cite{Joannopoulos-book}. Others are chiral metamaterials~\cite{chiral_Pendry}, transmission-line systems~\cite{crossing_transmission_lines}, bi-axial crystals~\cite{Berry} and spin systems~\cite{Rashba}. Previous studies of such systems have already shown some unusual properties~\cite{graphene,chiral_Pendry,Berry,segev_conical_diffraction}. However, these systems have been relatively unexplored, especially for pulse manipulations. For this purpose, the zero-gap system differs from a finite-gap system in the sense that while a pulse incident on a finite-gap system will mostly be reflected, a pulse incident on a zero-gap system will be almost perfectly admitted. 
Moreover, the proximity of the crossing bands allows for {\em efficient} transfer of energy between the bands using {\em weak} and {\em slow} modulations of the wave velocity. This enables light slowing-down, stopping or reversing etc..

In this Letter, we focus on time-reversal. A time-reversed pulse evolves as if time runs backward, thus eliminating any distortions or scattering that occurred at earlier times. It has applications in diverse fields such as medical ultrasound~\cite{review_Fink}, optical communications and adaptive optics~\cite{Pepper-book,Fink_Science,Fink_microwave_reversal}, superlensing~\cite{tr_super_lens_pendry}, ultrafast plasmonics~\cite{Stockman_reversal}, biological imaging~\cite{yaqoob_Nat_Phot}, THz imaging~\cite{THz_imaging} and quantum information and computing~\cite{quantum_reversal}. In particular, we study time-reversal of electromagnetic pulses which, to date, has been demonstrated for pulses of a relatively narrow spectrum or requires complicated schemes~\cite{Pepper-book,Yariv_Fekete_Pepper,Miller_OL,Marom_Fainman,Fan-reversal,Longhi-reversal,Kuzucu_Gaeta}. Some of these techniques~\cite{Fan-reversal,Longhi-reversal} are based on dynamically tuned PhCs. This novel subject of PhC research is experiencing a fast growth with applications such as frequency shifting, switching and many others studied theoretically and experimentally~\cite{Notomi_review,Lipson_review}. In this Letter, we show that dynamically tuned zero-gap PhCs enable an effective and broadband reversal in conceptually simple systems which are well within contemporary fabrication capabilities. Most importantly, unlike previous schemes, the zero-gap systems allow for reversal of pulses of unprecedented broad spectrum, thus, opening the way to reversal of few cycle pulses. In addition, as our scheme does not rely on any concept which is unique to optics, it can be applied for reversal of other wave systems for which time-reversal was not accessible before. 

\begin{figure}[htbp]
\centering{\includegraphics[scale=0.76]{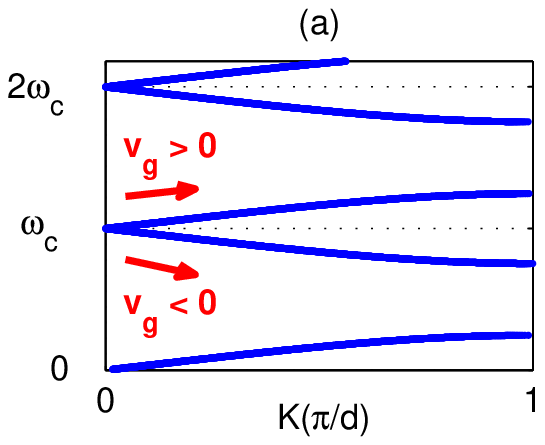} 
\includegraphics[scale=0.76]{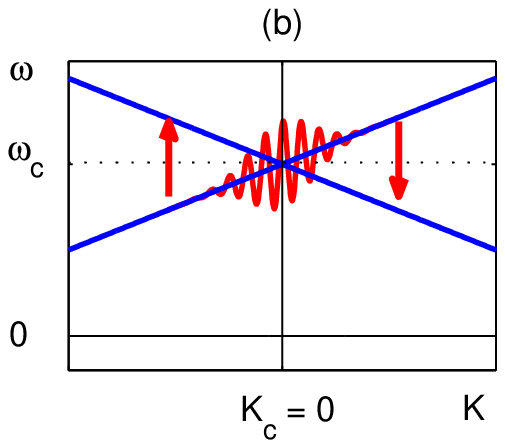} 
} \caption[]{(Color online) (a) Reduced Brillouin zone plot of the band structure of a QWS PhC. The bands support pulse propagation in opposite directions (indicated by the red arrows). (b) Schematics of the frequency conversion process. } \label{fig:qws_dispersion} 
\end{figure}

We consider the simplest example of a {\em perfectly symmetric zero-gap} system, namely, a 1D {\em layered} PhC satisfying the quarter-wave stack (QWS) condition, i.e., when the indices and thicknesses of the two layers satisfy~\cite{Joannopoulos-book,Yeh-book}
\begin{equation}\label{eq:qws_cond}
n_1 d_1 = n_2 d_2.
\end{equation}
Such a system has a zero-width gap at, e.g.,
\begin{equation}\label{eq:lambda_cr}
\lambda_c = 2 n_1 d_1 = 2 n_2 d_2. 
\end{equation}
The band structure of such PhCs can be computed analytically~\cite{Yeh-book}, see e.g. Fig.~\ref{fig:qws_dispersion}(a). At the crossing point (at zero Bloch momentum, $K = 0$), light can travel either in the forward or backward direction (i.e., on the positive/negative group-velocity band). An index modulation causes a transfer of energy between these bands. Intuitively, accurate reversal can be achieved only if two conditions are fulfilled. First, the modulation should be periodic in order to avoid any wavevector mixing, resulting in a vertical frequency-conversion, see Fig.~\ref{fig:qws_dispersion}(b). Second, the modulation should be much faster than the pulse duration or equivalently, the spectral content of the modulation should be much wider than that of the pulse; only such a {\em non-adiabatic} modulation ensures a frequency-independent frequency conversion, hence, an accurate reversal. Nevertheless, the modulation can still be much longer than the period of the pulse oscillations.

In what follows, we analyze the QWS system and confirm the intuitive arguments given above. However, we emphasize that the ideas and techniques we use here can be employed also in other zero-gap systems. 

\begin{figure}[htbp]
\centering{\includegraphics[scale=0.32]{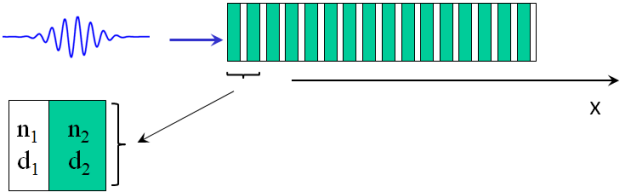}}
\caption[]{(Color online) Geometry of the pulse propagation through the QWS PhC. }	\label{fig:geometry}
\end{figure}

Consider an electromagnetic pulsed plane-wave normally incident on a QWS PhC (see Fig.~\ref{fig:geometry}) which is time-modulated in the following form
\begin{equation}\label{eq:n}
n(x,t) = n_{QWS}(x) + \Delta p(x)m(t), \quad max[m(t)] = 1. \nonumber
\end{equation}
Here, the static QWS $n_{QWS}(x)$ and spatial pattern of the modulation $p(x)$ have period $d$, and $\Delta$ is a constant representing the magnitude of the modulation. For a linear polarization, the Maxwell equations are given by
\begin{eqnarray}\label{eq:Maxwell2}
E_x(x,t) = - \mu_0 H_t(x,t), \ H_x = - \epsilon_0 \left[n^2(x,t) E(x,t)\right]_t, 
\end{eqnarray}
or, if the $H$-field is eliminated, by
\begin{eqnarray}\label{eq:waeq}
E_{xx}(x,t) = \left[v^{-2}(x,t) E(x,t)\right]_{tt}, 
\end{eqnarray}
which is a one-dimensional scalar wave equation with a space and time-dependent velocity $v = c/n(x,t)$.

We proceed with an analysis similar to that first performed in~\cite{deep_gratings_PRE_96}. We define 
\begin{eqnarray} \label{eq:EH_uni-dir}
E(x,t) &=& n^{-\frac{1}{2}}(x,t) \left[F(x,t) + B(x,t)\right], \nonumber \\
Z_0 H(x,t) &=& n^{\frac{1}{2}}(x,t) \left[F(x,t) - B(x,t)\right], \nonumber
\end{eqnarray}
where $F$ and $B$ are scalar functions representing the forward and backward fluxes in the PhC. 
Substituting in Eq.~(\ref{eq:Maxwell2}) gives
\begin{eqnarray}
F_x(x,t) + \frac{n(x,t)}{c} F_t + \frac{n_t}{c} F &=& - \frac{1}{2}\left(\frac{n_t}{c} - \frac{n_x}{n}\right) B, \nonumber \label{eq:F} \\
B_x(x,t) - \frac{n(x,t)}{c} B_t - \frac{n_t}{c} B &=& \frac{1}{2}\left(\frac{n_t}{c} + \frac{n_x}{n}\right) F. \nonumber \label{eq:B}
\end{eqnarray}
We rewrite the system in matrix form as
\begin{equation}\label{eq:W}
n\
\underline{W}_t =
\underline{\underline{M}}\ \underline{W},
\end{equation}
where
\begin{eqnarray} \label{eq:WM}  \nonumber
\underline{W} = \begin{pmatrix}
F \\
B
\end{pmatrix}, \quad \ \underline{\underline{M}} &=& \begin{pmatrix}
  - c \partial_x - n_t & - \frac{c}{2}\left(\frac{n_t}{c} - \frac{n_x}{n}\right) \\
 - \frac{c}{2}\left(\frac{n_t}{c} + \frac{n_x}{n}\right) & c \partial_x - n_t
\end{pmatrix}. \nonumber \label{eq:M}
\end{eqnarray}
Near a gap, the solution of Eq.~(\ref{eq:W}) consists predominantly of two spectral components on each of the two bands. We write each component as the product of a carrier uni-directional Floquet-Bloch (FB) mode $\Psi_{f/b}$ and a slowly varying envelope $f/b$, respectively, i.e.,
\begin{eqnarray}\label{eq:ansatz}
\underline{W} = \left[f(x_1,t_1) \underline{\Psi}_f(x_0) + b(x_1,t_1) \underline{\Psi}_b(x_0)\right] e^{- i \omega_c t_0} 
\end{eqnarray}
where $\omega_c = 2 \pi c / \lambda_c$ and $\underline{\Psi}_{f/b}$ are the eigenmodes of Eq.~(\ref{eq:W}). We also formally distinguish between fast variations of the carrier waves (through $x_0$ and $t_0$) and slower variations of the envelopes (through $x_1$ and $t_1$).

Assuming the central frequency of the input pulse is tuned exactly to the crossing point, substituting Eq.~(\ref{eq:ansatz}) in Eq.~(\ref{eq:W}), multiplying by $\Psi_f^*$ and $\Psi_b^*$, respectively, and integrating over $x_0$ allows us to remove the dependence on the fast scales. Then, for a piecewise-uniform modulation, we obtain the following equations for the time evolution of the forward/backward envelopes
\begin{eqnarray} \label{eq:f_uniform}
\left(1 + \Delta m^{(d)} m(t)\right) f_t(x,t) + v_g f_x + \Delta m^{(d)}\left(m_t - i \omega_c m(t)\right) f \nonumber \\
= \Delta m^{(od)} \left(i \omega_c m(t) b - m_t b - m(t) b_t\right), \nonumber \\
\end{eqnarray}
and
\begin{eqnarray} \label{eq:b_uniform}
\left(1 + \Delta m^{(d)} m(t)\right) b_t(x,t) - v_g b_x + \Delta m^{(d)}\left(m_t - i \omega_c m(t) \right) b \nonumber \\
= \Delta m^{(od)} \left(i \omega_c m(t) f - m_t f - m(t) f_t\right), \nonumber \\
\end{eqnarray}
where $v_g = c/\sqrt{n_1 n_2}$ is the group velocity and $m^{(d)} \equiv m^{ff} = m^{bb}$, $m^{(od)} \equiv m^{fb} = {m^{fb}}^*$ with
\begin{eqnarray} \nonumber
m^{ij} &=& \int_0^d \underline{\Psi}^\dagger_i(x) p(x) \underline{\Psi}_j(x) dx. \label{eq:m0_ab} 
\end{eqnarray}
For simplicity of notation, we have dropped the subscript of the slow scales coordinates.

In Eqs.~(\ref{eq:f_uniform})-(\ref{eq:b_uniform}), the terms proportional to $m(t)$ on the left-hand-sides describe the correction to the wave velocity, or equivalently, the correction to the phase accumulation, induced by the modulation. The terms proportional to $m_t(t)$ represent the change of the electromagnetic energy induced by the modulation. The terms on the right-hand-sides of Eqs.~(\ref{eq:f_uniform})-(\ref{eq:b_uniform}) are responsible for {\em reversal} coupling, i.e., to the vertical transitions $\omega + \delta \omega \to \omega - \delta \omega$, see Fig. 1b.
One can interpret the magnitude of the associated coefficients $m^{(od)}$ as the measure of the phase mismatch between the modes immediately above and below the crossing frequency. The coupling tends to zero as the index contrast approaches zero, as expected. In all other cases, the dependence of the coupling coefficient on the indices is non-trivial. 

Eqs.~(\ref{eq:f_uniform})-(\ref{eq:b_uniform}) can be solved analytically if we neglect all the coupling terms in Eq.~(\ref{eq:f_uniform}). This is justified as long as the backward component, which is absent prior to the modulation, is small compared with the forward component. Under this assumption of a {\em weak coupling}, the solution of Eq.~(\ref{eq:f_uniform}) can be shown to be
\begin{eqnarray}
f\left(x^{(f)},t\right) &=& e^{\Phi(t)} f_0\left(x^{(f)}\right), \label{eq:f_sol_full} \nonumber \\
\Phi(t) &=& \Delta m^{(d)} \int \frac{i \omega_c m(t') - m_t(t')}{1 + \Delta m^{(d)} m(t')}, \nonumber
\end{eqnarray}
where $x^{(f)} = x - v_g \int \frac{dt'}{1 + \Delta m^{(d)} m(t')}$ represents a frame moving with the forward pulse and $f_0$ is the envelope of the pulse traveling in the forward direction before the onset of the modulation. In a similar manner, it can be shown that the backward wave~(\ref{eq:b_uniform}) is given by
\begin{eqnarray}
b\left(x^{(b)},t\right) &=& \Delta m^{(od)} e^{\Phi(t)} \bar{b}\left(x^{(b)},t\right), \label{eq:b_sol_weak}
\end{eqnarray}
where
\begin{eqnarray}
\bar{b}\left(x^{(b)},t\right) = \int h(t') f_0\left(x^{(b)} - \int \frac{2 v_g dt''}{1 + \Delta m^{(d)} m(t'')}\right) dt', \label{eq:b_bar} \nonumber
\end{eqnarray}
and
\begin{equation}\label{eq:impulse_response}
h(t) = \frac{- i \omega_c m(t) + m_t(t)}{\left[1 + \Delta m^{(d)} m(t)\right]^2}, \nonumber
\end{equation}
where $x^{(b)} = x + v_g \int \frac{dt'}{1 + \Delta m^{(d)} m(t')}$ is a frame moving with the backward pulse; this shows that the wave-front has indeed been reversed. At times long after the modulation has ended, $b$ is effectively given by a convolution of the forward wave $f_0$ with $h(t)$, which thus plays the role of the impulse response of the system. Since the durations of $f_0$ and $h(t)$ are $T_{mod}$ and $T_p$, respectively, Eq.~(\ref{eq:b_sol_weak}) shows that accurate reversal can be achieved only in the {\em non-adiabatic} limit
\begin{equation}\label{eq:accurate_rev}
T_{mod} \ll T_p,
\end{equation}
in which case the impulse function is essentially a delta function. In order to see that more clearly, consider a Gaussian modulation and Gaussian pulse,
\begin{equation}\label{eq:gaussians}
m\left(t\right) = e^{- \frac{t^2}{T_{mod}^2}}, \quad \quad f_0(x - v_g t) = e^{- \frac{(x - v_g t)^2}{v_g^2 T_p^2}}.
\end{equation}
Then, for a small modulation ($\Delta \ll 1$), we get that $x^{(f,b)} = x \mp v_g t$, and the backward wave amplitude becomes
\begin{eqnarray}
b \cong - i \sqrt{\pi} \omega_c T_{eff} \Delta m^{(od)} e^{\Phi(t)} f_0\left(\frac{x + v_g t}{T_{eff}}\right),
\label{eq:b_sol_weak_small}
\end{eqnarray}
with $T_{eff} = T_{mod} T_p/\sqrt{4T_{mod}^2 + T_p^2}$. Thus, the reversal is accurate if condition~(\ref{eq:accurate_rev}) is satisfied; otherwise, the reversed pulse is broadener than the input pulse.

Eq.~(\ref{eq:b_sol_weak}) can also be solved asymptotically without any simplifying assumptions. However, whenever the backward wave becomes comparable in magnitude to the forward wave, the weak coupling solutions~(\ref{eq:b_sol_weak}) or~(\ref{eq:b_sol_weak_small}) are not valid anymore. In such cases, Eqs.~(\ref{eq:f_uniform})-(\ref{eq:b_uniform}) should ne solved {\em without} approximation.

We now turn to discuss some practical aspects of a Reversal Mirror (RM) design for electromagnetic pulses. First, the RM should be long enough to contain the pulse during the modulation. Simple estimates show that a few cm long RM can contain and reverse pulses not longer than a few tens of picoseconds. Second, the thicknesses of the layers determine the carrier (crossing) frequency according to Eq.~(\ref{eq:lambda_cr}). Third, with regards to modulation techniques, picosecond and longer modulation times can be achieved by exploiting the electro-optical effect or using carrier injection in semiconductors. In the latter case, index modulations of the order of $1\%$ at about $100$GHz~\cite{Lipson_review} or even at $1$ps-scale~\cite{ps_Si_modulation} were achieved in silicon devices. Shorter modulations must rely on an intense pulse in a pump-probe configuration, see e.g.,~\cite{Miller_OL}. In these cases, however, the reversal efficiency will be smaller due to the shorter modulation time, see Eq.~(\ref{eq:b_sol_weak_small}). 

\begin{figure}[htbp] 
\centering{\includegraphics[scale=0.95]{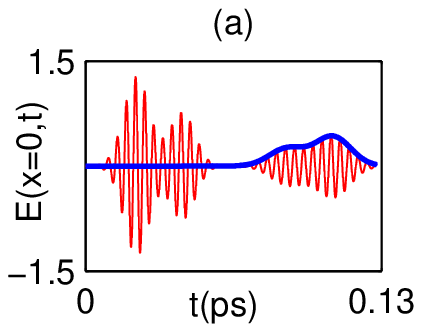} 
\includegraphics[scale=0.95]{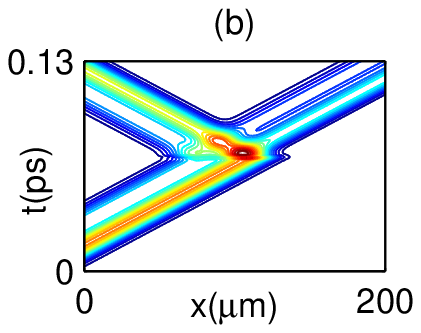}} 
\caption[]{(Color online) (a) Amplitude of an asymmetric pulse (Eq.~(\ref{eq:waeq}); red line) and the associated backward wave envelope (Eq.~(\ref{eq:b_uniform}); blue line) at the input side of the QWS as a function of time. (b) A spatio-temporal countour map of forward and backward envelopes in (a). } 	\label{fig:rev_illustration}
\end{figure}

\begin{figure}[htbp]
\centering{\includegraphics[scale=1.00]{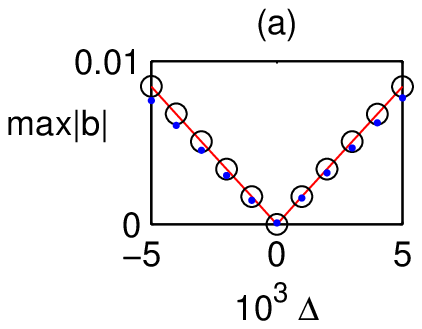} \includegraphics[scale=1.00]{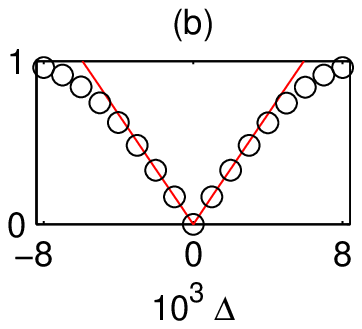}} 
\caption[]{(Color online) (a) Reversal of a $T_p = 100$fs, unit-amplitude Gaussian input pulse under a Gaussian modulation~(\ref{eq:gaussians}) as a function of the index change $\Delta$ for $n_1 = 3.45$, $n_2 = 1$, $\lambda_c = 1550$nm, and $T_{mod} = 10$fs. Shown are numerical solutions of the wave equation~(Eq.~(\ref{eq:waeq}); blue dots) vs. the solution of the envelope equations~(Eq.~(\ref{eq:b_uniform}); black circles) and the analytical solution~(Eq.~(\ref{eq:b_sol_weak_small}); red solid line). (b) Same as (a) for $T_{mod} = 1$ps and $T_p = 10$ps. } 	\label{fig:rev_efficiency}
\end{figure}

Finally, we demonstrate the performance of our RM through numerical simulations. In Fig.~\ref{fig:rev_illustration}(a), we plot the wave amplitude at the input side of the PhC as a function of time. The reversed pulse has a somewhat lower-amplitude, but the leading and trailing edges have clearly exchanged roles. We also show that the solutions of the wave equation~(\ref{eq:waeq}) and the envelope equations~(\ref{eq:f_uniform})-(\ref{eq:b_uniform}) are in excellent agreement. Fig.~\ref{fig:rev_illustration}(b) shows a spatio-temporal countour map of the pulse propagation. It shows that the modulation, occuring once the pulse is in the middle of the RM, causes a fast and complicated dynamics after which the pulse splits into a reversed and a (somewhat delayed) forward component.

A comparison of the reversal efficiencies as a function of the modulation strength, which also includes the analytical approximation~(\ref{eq:b_sol_weak_small}), is shown in Fig.~\ref{fig:rev_efficiency}. We employ realistic parameters corresponding to a silicon-air QWS with $\lambda_c = 1550$nm; such a PhC can be fabricated on-chip with current technology~\cite{Lipson_review}. First, we perform simulations of $\sim20$-cycle pulses. Fig.~\ref{fig:rev_efficiency}(a) verifies that all three solutions are in good agreement, thus, validating the analysis. Fig.~\ref{fig:rev_efficiency}(b) shows simulations for longer pulses for which memory and running time required for the solution of Eq.~(\ref{eq:waeq}) are very long. Thus, we only show the solutions of the envelope equations~(\ref{eq:f_uniform})-(\ref{eq:b_uniform}) vs. the analytical solution. As in Fig.~\ref{fig:rev_efficiency}(a), there is very good agreement between the numerical and analytical solutions up to high efficiencies ($\sim 50\%$). At even higher efficiencies, the analysis overestimates the reversal efficiency given by the solution of the envelope equations. This is because at such high efficiencies, the forward wave amplitude is significantly decreased, so the forward-to-backward wave coupling decreases as well. Nevertheless, the solutions of the envelope equations show that a $100\%$ reversal efficiency can be achieved with index modulations only slightly stronger than those predicted analytically. Thus, our RM has comparable performance to the previously suggested time-reversal schemes in index-modulated coupled-resonator arrays of optical waveguides~\cite{Fan-reversal,Longhi-reversal} and wave-mixing-based schemes~\cite{Miller_OL}. However, in contrast to these schemes, the QWS can admit and reverse pulses of very broad spectrum. In addition, our system does not suffer from a deterioration of performance due to non-adiabatic modulations.

In summary, we have shown how to time-reverse wave packets by exploiting the unusual band structure of zero-gap PhCs using index modulations that can be very weak and much slower than the wave period. The suggested design was shown to allow for a $100\%$ reversal efficiency, and can be applied to pulses of unprecedented broad spectrum. Our design can be fabricated with contemporary technology, e.g., on a silicon chip. Our RM does not require any prior knowledge on the pulse properties and can be implemented in almost any spectral range and pulse duration. Higher-dimensional realizations of zero-gap PhCs, e.g., having a hexagonal (graphene-like) symmetry, may open the way to additional reversal functionalities and may even be used as a building block in more sophisticated time-reversal schemes or other wave manipulation schemes like light stopping, slowing-down etc.. Finally, since our scheme does not rely on any concept which is unique to optics, it can be employed in other zero-gap systems for which time-reversal may not have been accessible so far, such as quantum systems~\cite{quantum_reversal}, spintronics~\cite{Rashba} and microwave transmission-lines~\cite{crossing_transmission_lines}.

Y.S. would like to thank A.I. Fernand\'ez-Domingu\'ez and A. Aubry for innumerable valuable discussions. Y.S. was supported by the Royal Society International Newton Fellowship. J.B.P was supported by the EU project PHOME (Contract No. 213390).

\end{document}